\documentclass[conference]{IEEEtran}
\IEEEoverridecommandlockouts
\usepackage{cite}
\usepackage{amsmath,amssymb,amsfonts}
\usepackage{graphicx}
\usepackage{textcomp}
\usepackage{booktabs}
\usepackage{algorithm}
\usepackage{algpseudocode}
\usepackage{xcolor}
\usepackage{float}
\usepackage{placeins}
\usepackage{tabularx}
\usepackage{multirow}
\renewcommand{\baselinestretch}{0.97}
\def\BibTeX{{\rm B\kern-.05em{\sc i\kern-.025em b}\kern-.08em
    T\kern-.1667em\lower.7ex\hbox{E}\kern-.125emX}}

\begin{document}

\title{Continuous Behavioral Authentication via Multi-Expert BERT Log Analysis for Secure Data Sharing\\}

\author{\IEEEauthorblockN{Stergios Lantzos, Ilias Syrigos, Apostolos Apostolaras, Thanasis Korakis}
\IEEEauthorblockA{\textit{Dept. of Electrical and Computer Engineering, University of Thessaly, Greece}}
\IEEEauthorblockA{\textit{Centre for Research and Technology Hellas, CERTH, Greece}}
\IEEEauthorblockA{lstergios@uth.gr, ilsirigo@uth.gr, apaposto@uth.gr, korakis@uth.gr}
}

\maketitle

\begin{abstract}
Continuous authentication for mobile and zero-trust systems requires nonintrusive evidence confirming the enrolled user-device context remains valid after initial login. This paper presents a BERT log analysis framework for continuous behavioral authentication using Android system logs. The proposed pipeline parses logcat streams into event templates and dynamic variables, pre-trains a domain-adapted BERT encoder on Android log syntax, and fine-tunes three expert models for network/device identity, battery-transition timing, and Wi-Fi topology. The expert confidence scores are fused through a log-space transformation and a 5-nearest-neighbor distance classifier to generate a normality score that is provided to a Policy Decision Point (PDP) for risk-aware access control. Experiments on normal traces, controlled anomaly injections, and benign Wi-Fi perturbations indicate that multi-expert BERT log analysis can detect semantic, battery-timing, and topology deviations in the evaluated setting while maintaining sub-1\% False Positive Rate (FPR). The results suggest that Android system logs are a practical sensor-free signal for continuous authentication and user-device context assurance.
\end{abstract}


\section{Introduction}

Modern mobile devices have evolved into an integral part of daily life, handling sensitive and personal data facilitating personal communications, processing banking transactions, and storing personal archiving. At the same time, with the rapid growth of the data economy, mobile devices have become versatile ubiquitous platforms for participation in data-sharing infrastructures governed by zero-trust access policies. Consequently, authenticating and continuously verifying that legitimate device owners remain in control throughout data-sharing sessions is a fundamental pillar in modern authentication and zero-trust frameworks to ensure authorized control.

Traditional authentication methods rely on an entry-point authentication model, in which users are required to verify their identity at the start of the session using a username and password, a PIN, or biometric authentication. However, no matter how robust these mechanisms are for authenticating users, an attacker who gains physical access, after initial authentication is completed, can potentially exploit the static nature of those mechanisms and obtain afterwards sensitive information. Frequently re-prompting users for authentication is not a viable solution, as it degrades user experience and, as indicated by a comprehensive survey~\cite{Crawford2014TransparentAuth}, users prefer transparent authentication methods that are nonintrusive. Thus, continuous authentication schemes have emerged that verify the identity of users continuously in the background while the mobile device is in use. Such methods are often based on behavioral biometrics, including motion and gait, typing cadence and touchscreen dynamics, gestures, and voiceprints. Additionally, several works~\cite{Murmuria2015CA, Neal2015BTAS, NealWoodard2017Assoc, Jakubeit2022WiFiAuth} have demonstrated the value of interactive behavioral features such as application usage, power consumption, network identity and location fingerprints that do not require extraction of data from the device's sensors, while remaining user-specific and stable.

In this context, Android system logs represent a largely unexploited behavioral signal. Although prior work has shown that application usage sequences, Bluetooth device sightings, and Wi-Fi access-point associations are reliable contextual behavioral biometrics, they do not exploit the semantic structure of the log stream itself, such as variable co-occurrence rules and temporal hardware state transitions that the device continuously records at the operating-system level. On the log-analytics side, log anomaly detectors such as (BERT)-based LogBERT~\cite{Guo2021LogBERT} and others~\cite{Huang2023HilBERT, Xiao2023Loader, Chai2024LogSer, Xie2022LogGD} have demonstrated that Transformer language models can learn the ``grammar'' of system log sequences with high fidelity. These methods mainly target service reliability on server infrastructure, i.e., detecting software faults in Hadoop Distributed File System (HDFS) or BlueGene/L (BGL) logs and they produce a binary normal/anomalous verdict per log window. They are usually stateless across windows, agnostic to variable semantics, and are not designed to produce a continuous behavioral trust score for a specific physical user.

This paper lies at the intersection between the two lines of work: continuous user and device authentication and transformer-based log analytics. We observe that Android system logs simultaneously encode three independent behavioral dimensions: (i)~the network \emph{identity} of the device including the network entities with which the device interacts (MAC addresses, IP assignments, and Bluetooth peers), (ii)~the \emph{temporal dynamics} of hardware power consumption (battery charge/discharge intervals as a physical fingerprint of the owner's usage habits), and (iii)~the \emph{spatial topology} of the surrounding Wi-Fi environment. Fusing these three signals into a unified behavioral trust score that can be directly consumed by a Policy Decision Point (PDP) in a zero-trust architecture is the main problem addressed in this paper.

The contributions of this paper are:
\begin{itemize}
    \item A continuous behavioral authentication framework that exploits Android system logs as a nonintrusive source of user and device behavioral evidence.
    \item A log parsing pipeline and a BERT-based pre-training strategy for learning the structural grammar of Android system logs.
    \item A multi-expert architecture that separately models network/device identity, battery consumption dynamics, and Wi-Fi-based spatial topology through BERT-based models.
    \item A multi-expert fusion classifier that fuses the expert outputs into a trust score to be provided as input to a PDP in a zero-trust architecture.
    \item An evaluation of the framework under semantic, battery-consumption, and spatial anomaly scenarios.
\end{itemize}

The remainder covers related work (\S\ref{sec:related}), system architecture (\S\ref{sec:architecture}), data collection (\S\ref{sec:data}), pre-training and fine-tuning (\S\ref{sec:pretrain}--\S\ref{sec:finetune}), the fusion classifier (\S\ref{sec:knn}), evaluation (\S\ref{sec:eval}), and conclusions (\S\ref{sec:conclusion}).

\section{Related Work} \label{sec:related}

\subsection{Continuous Authentication and Behavioral Biometrics}

Physiological biometrics (fingerprint, face, iris) are susceptible to presentation attacks and, once compromised, cannot be revoked. In addition, they are inherently entry-point mechanisms that provide no guarantee of continued legitimate control after the initial check. Surveys by Abuhamad et al.~\cite{Abuhamad2020Survey} (140+ approaches, six modality groups) and Liang et al.~\cite{Liang2020IoTBiometrics} (IoT-wide AI perspective) establish that behavioral biometrics overcome these limitations by enabling transparent, nonintrusive continuous authentication (CA) and that 90\% of users prefer such passive verification over explicit credential prompts. Focusing on device-native behavioral signals, Murmuria et al.~\cite{Murmuria2015CA} showed that Android power consumption, touch, gestures, and movement form a viable ensemble (Equal Error Rate (EER)~6.1\%--6.9\%, 59~users). Neal et al.~\cite{Neal2015BTAS} demonstrated that application traffic, Bluetooth sightings, and Wi-Fi associations are distinctive fingerprints stable over 19 months (up to 93\% identification for Wi-Fi), while a companion study~\cite{NealWoodard2017Assoc} using association rule mining reached 91\% verification accuracy. Jakubeit et al.~\cite{Jakubeit2022WiFiAuth} were able to authenticate a device's registered environment against 11 other locations, achieving precision~$>$~0.98 and recall~$>$~0.99 across all test sites, from passive Wi-Fi beacon frames alone, without storing any personally identifiable information.

\subsection{Log Parsing and Anomaly Detection}

Log-based anomaly detection begins with log parsing, which is the transformation of raw, semi-structured messages into structured event templates and dynamic variable fields. This step is not trivial at production scale: Drain~\cite{He2017Drain}, an online prefix-tree parser, remains the dominant choice due to its efficiency and competitive accuracy, yet a large-scale re-evaluation by Jiang et al.~\cite{Jiang2024LogParsing} shows that all 15 surveyed parsers degrade substantially on infrequent and parameter-intensive templates, thus motivating the domain-specific adaptations we apply to Android system logs. LogBERT~\cite{Guo2021LogBERT} established the dominant paradigm for self-supervised log anomaly detection by combining masked log-key prediction with hypersphere minimization. Subsequent work extended this baseline in orthogonal directions: HilBERT~\cite{Huang2023HilBERT} adds a hierarchical encoder to handle log instability, Loader~\cite{Xiao2023Loader} replaces the Recurrent Neural Network (RNN) backbone with a Transformer and a confidence-adaptive top-$p$ rule, and Yan et al.~\cite{Yan2024Transformer} exploit unlabeled production logs via masked-reconstruction pre-training. LogSer~\cite{Chai2024LogSer} is closest to our work: it introduces named semantic tokens and fuses template with parameter embeddings, recognizing that dynamic variable values carry independent anomaly signals. LogGD~\cite{Xie2022LogGD} captures co-occurrence structure via a Graph Transformer.

\subsection{Research Gap}

All log anomaly detectors above target server infrastructure, produce stateless binary verdicts per window and they cannot provide a continuous trust score bound to a specific physical user.
CA systems, conversely, exploit Bluetooth, Wi-Fi, and power consumption as pre-aggregated statistics and never mine the raw OS-level log stream that natively encodes all three modalities: network identity, power consumption dynamics, and spatial topology. To the best of our knowledge, this paper is the first attempt to apply domain-decomposed BERT-based expert models to on-device Android logs for continuous behavioral authentication, fusing the aforementioned modalities into a PDP-ready trust score for zero-trust architectures.

\section{System Architecture} \label{sec:architecture}

\begin{figure*}[!t]
\centering
\includegraphics[width=0.91\textwidth]{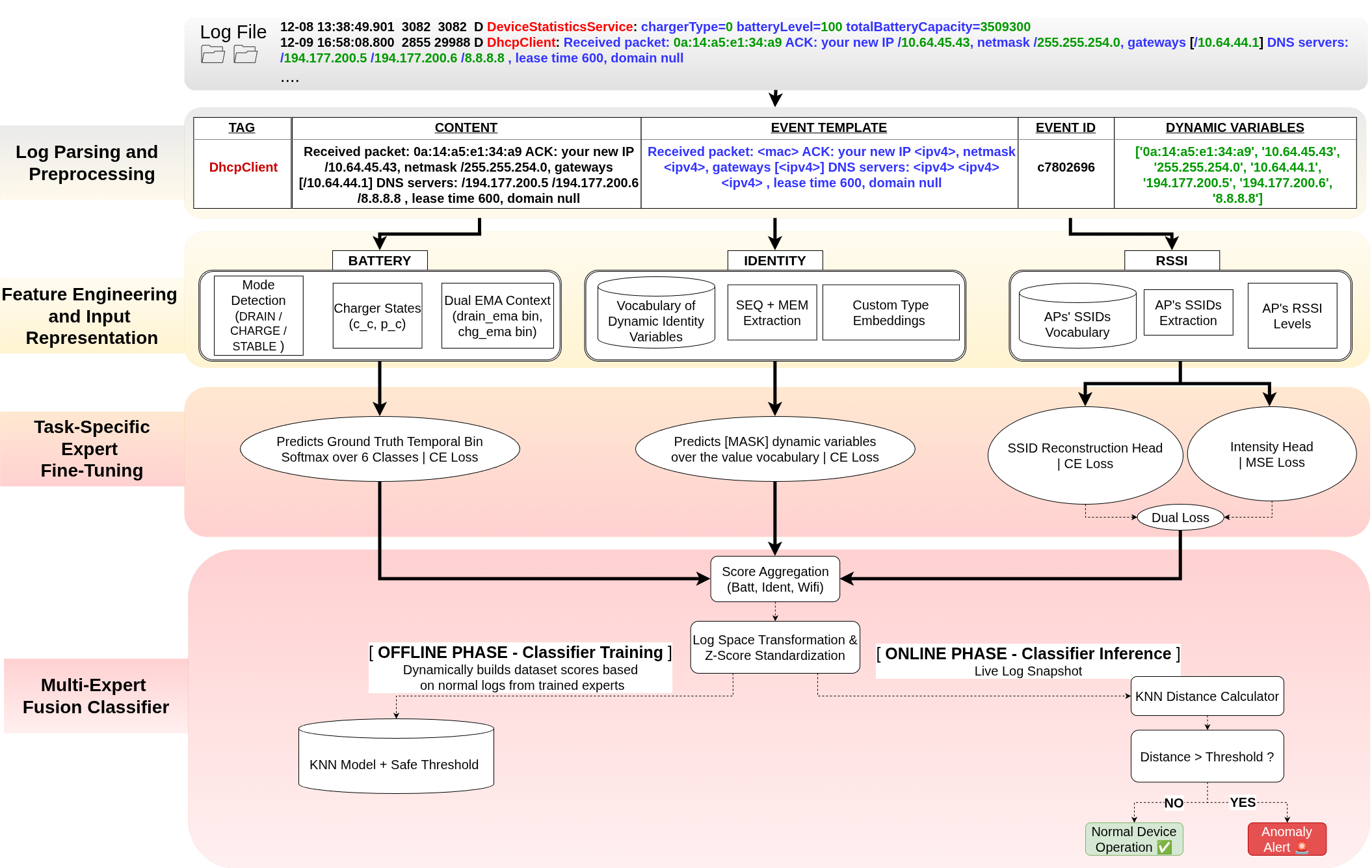}
\caption{Overview of the proposed continuous behavioral authentication framework.}
\label{fig:architecture}
\end{figure*}

Fig.~\ref{fig:architecture} illustrates the end-to-end pipeline of the proposed framework, which comprises four sequential phases.

\textbf{Phase~1: Log Parsing.}
First, raw Android system log files are ingested by a domain-adapted Drain parser~\cite{He2017Drain} (detailed in Section~\ref{sec:data}). Each log line is decomposed into a log tag, an event template with typed placeholder tokens (\texttt{<mac>}, \texttt{<ipv4>}, \texttt{<int>}, etc.), a stable event identifier derived from a hash of the template, and a list of extracted dynamic variable values.

\textbf{Phase~2: Domain-Specific Feature Engineering.}
The parsed output is routed into three semantically independent channels. The \emph{Identity} channel pairs a sliding window of the 50 most recent event IDs with the last observed value per event ID, injecting type-aware embeddings to encode variable category. The \emph{Battery} channel processes \textit{DeviceStatisticsService} logs via dual exponential moving-average (EMA) telemetry, encoding each snapshot as a two-segment sequence of operating mode and smoothed measurements. The \emph{Wi-Fi Topology} channel encodes \textit{WifiPickerTracker} scans as fused SSID--RSSI-level token pairs, one per visible access point.

\textbf{Phase~3: Expert Fine-Tuning.}
Each channel is processed by a BERT-based expert (Sections~\ref{sec:pretrain}--\ref{sec:finetune}). The \emph{Identity Expert} predicts masked dynamic variable values from a training-constrained vocabulary, where out-of-vocabulary (OOV) values naturally receive very low model confidence. The \emph{Battery Expert} classifies each snapshot into one of six temporal bins, making it invariant to absolute battery level. The \emph{Wi-Fi Topology Expert} uses dual heads to reconstruct SSID identity (Cross-Entropy) and RSSI levels using mean-squared error (MSE), jointly modeling the spatial topology.

\textbf{Phase~4: Multi-Expert Fusion Classifier.}
At inference time, each expert produces a per-snapshot confidence score (Section~\ref{sec:knn}). The three scores are concatenated and passed through a log-space scaling transform that suppresses intra-class variance while magnifying outlier distances. During an offline training phase, the transformed score vectors of normal-labeled snapshots are stored to build a \emph{normal region}. At runtime, the incoming transformed vector is compared against this region using a 5-nearest-neighbor distance-based classifier. An anomaly alert is raised when the resulting distance exceeds the 99.9th-percentile threshold derived during offline calibration, otherwise the snapshot is characterized as normal device operation.

\section{Data Collection and Log Preprocessing} \label{sec:data}

Logs were collected on a OnePlus~8T smartphone running Android~11.0 over a one-week period of normal daily use, yielding 13{,}058 parsed log entries for training and calibration across six collection sessions. The same device was also used to generate the controlled resource-exhaustion battery anomaly dataset described in Section~\ref{sec:eval}. Log acquisition relies on Android's \texttt{logcat} interface via a persistent background service.

Logs are collected from four interfaces (Table~\ref{tab:LogInterfaceDetails}), routed into three expert channels. The \emph{Identity} channel consolidates Wi-Fi and Bluetooth logs, recording network associations, IP/MAC assignments, and peripheral connections to form a behavioral fingerprint of the user's digital environment. The \emph{Battery} channel tracks power-subsystem transitions at 1\% granularity, encoding the user's physical usage rhythm. The \emph{Wi-Fi Topology} channel captures ambient Wi-Fi topology from periodic \textit{WifiPickerTracker} scans, providing a spatial fingerprint of the device's location.

\begin{table}[h!]
\centering
\caption{Monitored Android Log Tags and Functional Descriptions}
\label{tab:LogInterfaceDetails}
\resizebox{\columnwidth}{!}{%
\begin{tabular}{|l|l|l|}
\hline
\textbf{Interface} & \textbf{Log Tag} & \textbf{Operational Significance} \\ \hline
\textbf{WiFi} & IpClient & Monitors IPv6 route/address additions and removals. \\ \cline{2-3}
 & WifiClientModeImpl & Tracks gateway MAC address and device identity. \\ \cline{2-3}
 & DhcpClient & Manages IPv4 assignment, renewal, and server handshakes. \\ \cline{2-3}
 & ConnectivityService & Detects network state transitions (e.g., Connected to Disconnected). \\ \cline{2-3}
 & OpIpConflictDetector & Identifies ARP-based IP conflicts and gateway integrity. \\ \hline
\textbf{Bluetooth} & BluetoothManagerService & Monitors BLE state transitions (ON/OFF/Turning OFF). \\ \cline{2-3}
 & BluetoothRemoteDevices & Records remote device names, connections, and battery levels. \\ \cline{2-3}
 & CachedBluetoothDevice & Manages pairing status and bond removal commands. \\ \hline
\textbf{Battery} & DeviceStatisticsService & Logs battery level, capacity, and charger type (USB/AC/None). \\ \hline
\textbf{RSSI} & WifiPickerTracker & Provides real-time signal strength (RSSI) of nearby Access Points. \\ \hline
\end{tabular}%
}
\end{table}

\subsection{Domain-Adapted Drain Parser}

Raw log lines are heterogeneous semi-structured text. We adapt Drain~\cite{He2017Drain}, an online prefix-tree parser, to Android~11.0 by injecting domain-specific regular expressions that atomically identify MAC addresses, IPv4/IPv6 identifiers, BLE state strings, and integer fields before the tree traversal begins. Each parsed line is decomposed into: (i)~a stable \emph{event template} with typed placeholder tokens (\texttt{<mac>}, \texttt{<ipv4>}, \texttt{<ipv6>}, \texttt{<int>}, \texttt{ <string>}, \texttt{<bool>}), (ii)~a hash-derived \emph{event ID}, and (iii)~an ordered list of \emph{dynamic variable values}. Templates define the structural grammar the pre-trained BERT encoder learns in Section~\ref{sec:pretrain}, while dynamic values form the semantic vocabulary the expert models validate in Section~\ref{sec:finetune}.

\section{Pre-Training: Learning the Log Syntax} \label{sec:pretrain}

All three experts share a BERT-base backbone (12 encoder layers, 12 self-attention heads, 768-dimensional hidden state). During fine-tuning, the first 2 layers are frozen and the remaining 10 are task-adapted (Section~\ref{sec:finetune}). The tokenizer is extended with the six typed placeholder tokens from Section~\ref{sec:data} (\texttt{<mac>}, \texttt{<ipv4>}, \texttt{<ipv6>}, \texttt{<int>}, \texttt{<string>}, \texttt{<bool>}) to preserve semantic category throughout encoding.

Pre-training uses a domain-specific MLM variant: placeholder tokens are masked and the model predicts the correct placeholder \emph{type}, teaching it the co-occurrence grammar of the log stream. This structural knowledge is intentionally ``blind'' to actual variable \emph{values} (the encoder learns \emph{where} a \texttt{<mac>} appears, not which address is expected) so each expert fine-tuning stage need only learn user-specific behavioral patterns rather than log syntax from scratch.

\section{Expert Model Fine-Tuning} \label{sec:finetune}

Following pre-training, specialized prediction heads are attached to the shared BERT encoder, one per behavioral channel. Each expert is fine-tuned on user-specific data to learn the \emph{expected} distribution of values for its channel, transitioning the structurally-aware encoder into a per-user behavioral authenticator.

\subsection{Identity Expert}

The Identity Expert transitions the encoder from syntactic awareness, knowing \emph{where} a variable type appears, to semantic validation: knowing \emph{which specific value} is expected given the device's behavioral history.

\textbf{Input and Tokenization.}
The input uses \texttt{[CLS] SEQ [SEP] MEM [SEP] TASK + TARGET TEMPLATE}. \texttt{SEQ} contains only the last 50 Identity Event IDs, without dynamic variables, \texttt{MEM} stores the dynamic variables from the previous occurrence of the Event ID under evaluation, \texttt{TASK} is that Event ID, and \texttt{TARGET TEMPLATE} masks its dynamic-variable positions (e.g., \texttt{Unicasting DHCPREQUEST ciaddr=[MASK] request=[MASK] serverid=[MASK] to [MASK]:[MASK]}). Variables are omitted from \texttt{SEQ} because full parameterized windows often exceed BERT's 512-token limit and would propagate injected or swapped values through the next 50 contexts, increasing false positives. All unique training-set dynamic values are added as atomic tokens (Custom Dynamic Vocabulary) to prevent WordPiece fragmentation of structured identifiers (e.g., \texttt{194.177.200.5} split into seven tokens). For each masked variable, the model also receives an embedding of its placeholder type (e.g., MAC, IPv4, integer), so that predictions are conditioned on both context and expected value class:
\begin{equation}
    E_{final} = E_{token} + E_{position} + E_{type}
\end{equation}
where $E_{type}$ encodes the variable category and limits candidate values to the matching semantic class. OOV values are mapped to \texttt{[UNK]}. As this token is not observed in normal user-specific training contexts, the model assigns it very low confidence during evaluation, producing a low $S_{Ident}$ for unseen identity values.

\textbf{Training.}
The first two encoder layers are frozen to preserve the structural features acquired during pre-training. The remaining 10~layers and a classification head are trained with Cross-Entropy over the custom value vocabulary, while static tokens are excluded from the objective.

\textbf{Confidence Score.}
The per-log confidence score $S_{Ident} \in [0,1]$ is the arithmetic mean of the predicted Softmax probabilities at each dynamic variable position:
\begin{equation}
    S_{Ident} = \frac{1}{N} \sum_{j=1}^{N} p\!\left(\hat{v}_j = v_j \mid \mathbf{C}_{Ident}\right)
\end{equation}
where $N$ is the number of dynamic variable positions in the log, $v_j$ is the observed value, and $\mathbf{C}_{Ident}$ denotes the Identity Expert input context: the 50-event-ID \texttt{SEQ}, \texttt{MEM}, \texttt{TASK}, and masked target template.
\subsection{Battery Expert}

The Battery Expert models the temporal and logical behavior of the device’s power subsystem using two complementary abstractions: battery-transition states and temporal classes. First, each pair of consecutive battery observations is assigned to one of three transition states according to the direction of the battery-level change: \texttt{STABLE}, when the battery percentage remains unchanged, \texttt{DRAIN}, when the level decreases by 1\%, and \texttt{CHARGE}, when the level increases by 1\%. Second, for physically valid transitions, the elapsed time between observations is quantized into one of five temporal classes derived from K-Means clustering on the normal training data (Table~\ref{tab:TemporalCentroids}): \texttt{TS\_INSTANT}, \texttt{TS\_SHORT}, \texttt{TS\_MID}, \texttt{TS\_LONG}, and \texttt{TS\_SUSTAINED}. These classes capture the device’s normal power-consumption rhythm at different time scales, from short reporting intervals to sustained battery states. In addition, a separate rule-based class, \texttt{TS\_VIOLATION}, captures physically inconsistent transitions, such as a discharging event while the device is reported as connected to AC power. The Battery Expert therefore learns not only how quickly battery transitions normally occur, but also whether the observed transition is logically consistent with the device’s charger state.

\textbf{Level-Agnostic Input.}
To eliminate dependency on absolute battery level (the ``Level Trap''), the gauge value is omitted from the input. Each snapshot is encoded as a two-segment BERT sequence: Segment~A carries the operational mode (\texttt{STABLE}/\texttt{DRAIN}/\texttt{CHARGE}) and Segment~B carries the telemetry context:
\begin{equation}
\resizebox{\columnwidth}{!}{%
$\texttt{[CLS]}\ \underbrace{\texttt{Mode}}_{\text{Seg.~A}}\ \texttt{[SEP]}\
\underbrace{\texttt{c\_c=[C\_C]\ p\_c=[P\_C]\ drain\_ema=[D\_EMA]\
chg\_ema=[C\_EMA]}}_{\text{Seg.~B}}\ \texttt{[SEP]}$%
}
\end{equation}
where \texttt{c\_c} and \texttt{p\_c} denote the current and previous charger states (USB/AC/None), and \texttt{drain\_ema} and \texttt{chg\_ema} are EMA-smoothed estimates ($\alpha=0.6$) of the battery percentage change per unit time during drain and charge phases, updated recursively as $R_{EMA}^{(t)} = \alpha \cdot R_{current} + (1-\alpha) \cdot R_{EMA}^{(t-1)}$. The final input representation is:
\begin{equation}
    E_{input} = E_{token} + E_{position} + E_{segment}
\end{equation}
where $E_{segment}$ trains the attention heads to cross-reference the operational mode against telemetry, enabling detection of cross-segment contradictions such as a \texttt{DRAIN} state with an active AC connection, or a \texttt{CHARGE} state with no charger reported.

\textbf{Training.}
The Battery Expert is trained as a Multi-Class Classifier over the 6 bins using Categorical Cross-Entropy with normalized soft labels. For valid temporal ground-truth bin $c^*$ and neighboring-bin set $\mathcal{N}(c^*)$, the target distribution is:
\begin{equation}
    Z = 0.8 + 0.1|\mathcal{N}(c^*)|
\end{equation}
\begin{equation}
    y'_{c} =
    \begin{cases}
        0.8/Z & c = c^* \\
        0.1/Z & c \in \mathcal{N}(c^*) \\
        0     & \text{otherwise.}
    \end{cases}
\end{equation}
Thus middle bins keep the usual $0.1/0.8/0.1$ target, while edge bins such as \texttt{TS\_INSTANT} or \texttt{TS\_SUSTAINED} receive $0.889$ for the ground truth and $0.111$ for their single neighbor. Synthetic \texttt{TS\_VIOLATION} samples are injected as hard targets ($y'=1.0$).

\textbf{Confidence Score.}
The per-snapshot confidence score $S_{Batt} \in [0,1]$ combines the Softmax probability assigned to the observed valid temporal bin $c_{\mathrm{obs}}$ with the probability assigned to the logical-violation class $p_{viol}$:
\begin{equation}
    S_{Batt} = p_{c_{\mathrm{obs}}}(1 - p_{viol}).
\end{equation}
A high value indicates that the observed inter-event interval and charger state are consistent with the model's learned discharge rhythm, whereas a low value signals a timing or logical-consistency anomaly. $S_{Batt}$ is propagated to the Multi-Expert Fusion Classifier, where the nearest-neighbor distance test evaluates it jointly with the Identity and Wi-Fi Topology scores. 

\begin{table}[h!]
\centering
\caption{Temporal Binning: K-Means Centroids and Decision Boundaries}
\label{tab:TemporalCentroids}
\begin{tabular}{@{}llll@{}}
\toprule
\textbf{Bin} & \textbf{Centroid (s)} & \textbf{Range (min)} & \textbf{Significance} \\ \midrule
\texttt{TS\_INSTANT}   & 30    & 0.0\,--\,1.75   & Immediate State  \\
\texttt{TS\_SHORT}     & 180   & 1.75\,--\,9.0   & Short Transition \\
\texttt{TS\_MID}       & 900   & 9.0\,--\,28.75  & Standard State   \\
\texttt{TS\_LONG}      & 2550  & 28.75\,--\,58.75& Extended State   \\
\texttt{TS\_SUSTAINED} & 4500  & $>$\,58.75      & Sustained State  \\
\texttt{TS\_VIOLATION} & ---   & ---             & Logic Violation  \\ \bottomrule
\end{tabular}
\end{table}

\subsection{Wi-Fi Topology Expert}

The Wi-Fi Topology Expert is trained as a dual-head autoencoder to reconstruct the device's Wi-Fi topology from \textit{WifiPickerTracker} scans. As with the Identity Expert, SSIDs are added as atomic tokens to a Custom SSID Vocabulary. Unknown SSIDs at inference time are mapped to \texttt{[UNK]}, indicating a potential out-of-distribution environment. Each access point is represented as a fused identity--intensity embedding:
\begin{equation}
    E_{final} = [E_{SSID} + E_{Level}] + E_{position}
\end{equation}
allowing the self-attention mechanism to learn inter-AP spatial correlations (e.g., ``if AP-A is at level~4, AP-B is expected at level~2'').

\textbf{Training.}
The dual-head objective separates the two components of the Wi-Fi topology: access-point identity and signal-strength intensity. The SSID head is trained with Cross-Entropy over the categorical SSID vocabulary, while the intensity head is trained with Mean Squared Error (MSE) over the ordinal RSSI level representation. The total reconstruction objective is defined as a weighted multi-task objective:
\begin{equation}
L_{\mathrm{Total}} =
\lambda_{\mathrm{SSID}} L_{\mathrm{SSID}} +
\lambda_{\mathrm{Level}} L_{\mathrm{Level}},
\end{equation}
where $\lambda_{\mathrm{SSID}}$ and $\lambda_{\mathrm{Level}}$ are nonnegative weighting coefficients that control the relative contribution of the identity and intensity reconstruction terms. In our experiments, both are set to 1, yielding an equal-weight objective. Under this formulation, unfamiliar or incorrectly reconstructed SSIDs increase $L_{\mathrm{SSID}}$, while deviations in the expected signal-strength levels increase $L_{\mathrm{Level}}$, so the total objective captures both semantic and spatial deviations from the learned Wi-Fi topology.

\textbf{Confidence Score.}
The per-scan confidence score $S_{WiFi} \in [0,1]$ combines the outputs of the SSID and intensity heads. For each visible access point $k$, the SSID head assigns a Softmax probability $p_k$ to the observed SSID, while the intensity head produces a reconstruction error $\ell_k$ for the corresponding RSSI level. The scan-level score is computed as:
\begin{equation}
S_{WiFi} =
\frac{1}{M}
\sum_{k=1}^{M}
p_k \cdot e^{-\ell_k},
\end{equation}
where $M$ is the number of visible access points. Unknown SSIDs receive near-zero identity probability, substantially reducing the score in an unfamiliar environment. Absent access points are excluded from the sum, so partial but spatially consistent topologies retain high confidence.

\section{Multi-Expert Fusion Classifier} \label{sec:knn}

Each expert independently assesses one behavioral dimension of the enrolled user-device context and emits a scalar confidence score $S \in [0,1]$. The framework therefore authenticates continuity of the enrolled user-device behavioral context rather than relying on a single physiological biometric. While a sufficiently strong deviation in any single channel triggers an alert, realistic attacks may be crafted to remain below the detection threshold of every individual expert. The Multi-Expert Fusion Classifier addresses this by treating the three scores as a joint behavioral fingerprint and evaluating their combination against a manifold of learned normal behavior. Its design follows three sequential steps: a nonlinear feature transformation to amplify sensitivity near the high-confidence region, a Z-score standardization to equalize inter-expert variance, and a nearest-neighbor distance test against a stored normal region.

\subsection{Expert Score Aggregation and Log-Space Transformation}

At every inference step, the system maintains a persistent snapshot of the device's behavioral state as a three-dimensional score vector:
\begin{equation}
    \mathbf{v} = [S_{Ident},\ S_{Batt},\ S_{WiFi}], \quad S_i \in [0,1]
\end{equation}
where $S_{Ident}$, $S_{Batt}$, and $S_{WiFi}$ are the confidence scores emitted by the Identity, Battery, and Wi-Fi Topology experts, respectively. Raw confidence scores are poorly suited for direct distance comparison: normal behavior clusters near $S \approx 1.0$, so small deviations from normality produce imperceptibly small differences in Euclidean space. To address this, we apply a nonlinear log-space transformation:
\begin{equation}
    X_{log} = -\ln(S + \epsilon)
\end{equation}
where $\epsilon = 10^{-15}$ prevents numerical collapse at $S = 0$. This negative-log structure maps perfect confidence to zero and strongly expands low-confidence scores, mapping $[0,1]$ onto an unbounded range where anomalous scores are geometrically separated from the normal cluster.

\subsection{Z-Score Standardization and Normal Region Construction}

Because the three experts operate under different objectives and output scales, their raw log-transformed scores exhibit different variances. To ensure that each dimension contributes equally to the distance metric, the transformed scores are standardized using the mean $\mu_i$ and standard deviation $\sigma_i$ computed exclusively from the normal training set:
\begin{equation}
    z_i = \frac{X_{log,i} - \mu_i}{\sigma_i}
\end{equation}
This maps the normal training distribution onto a 3D \emph{Normal Region} centered at the origin, where deviations along each axis reflect proportional anomalousness in the corresponding expert channel. The complete set of transformed normal-training vectors is stored as the reference structure for the distance classifier.

\begin{figure}[h!]
    \centering
    \includegraphics[width=0.9\columnwidth]{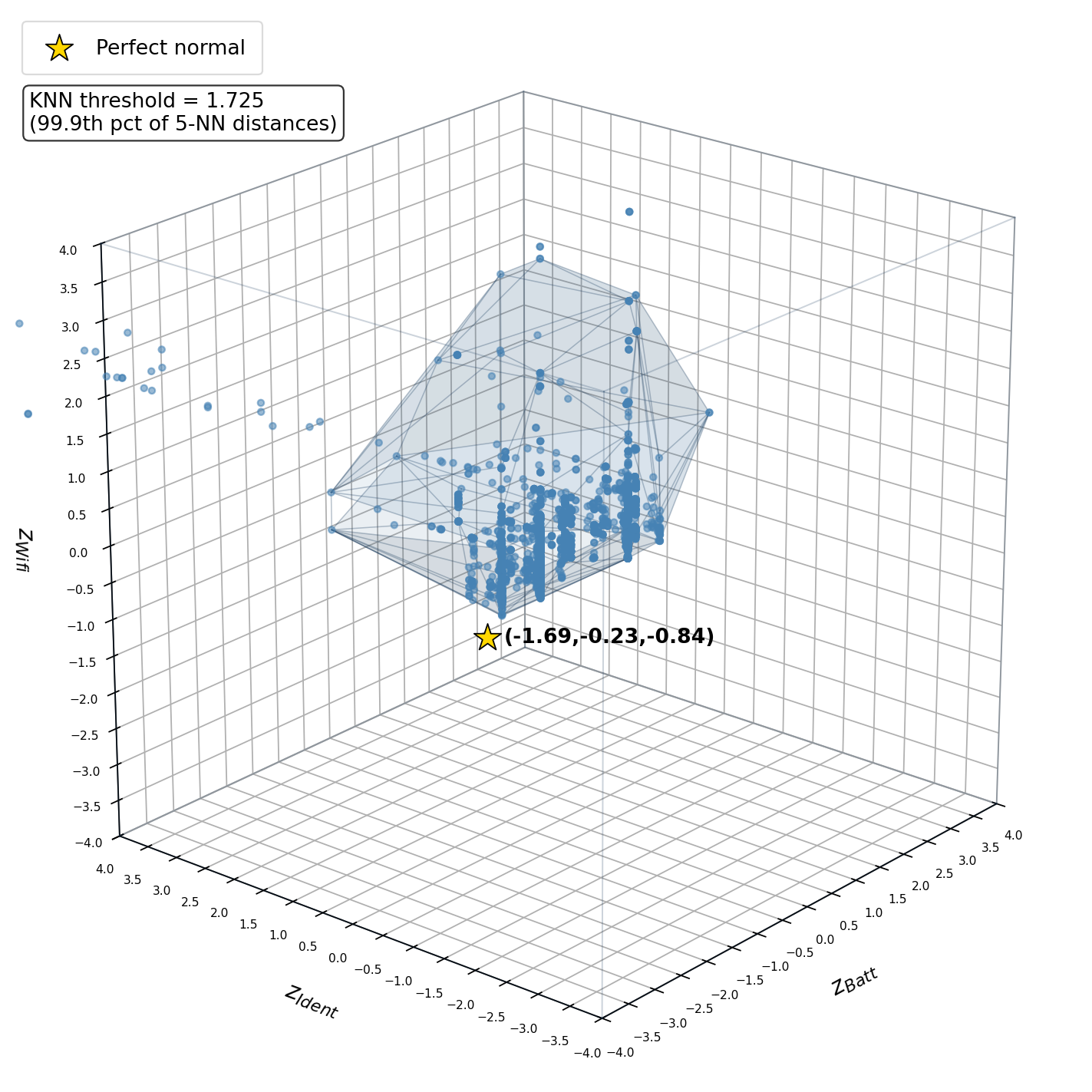}
    \caption{Z-score normal region. Axes show standardized expert log-costs, while the convex hull encloses normal behavior and the 99.9th-percentile 5-NN threshold $\tau=1.725$ defines the anomaly boundary.}
    \label{fig:zscore_manifold}
\end{figure}

\subsection{Nearest-Neighbor Distance Classifier}

We adopt an unsupervised 5-nearest-neighbor distance classifier to define the boundary of normal behavior, trained exclusively on clean historical data. At inference time, the classifier identifies the set $\mathcal{N}_5(\mathbf{z})$ of the five nearest stored normal vectors and computes the mean Euclidean distance:
\begin{equation}
    D(\mathbf{z}) = \frac{1}{5}
    \sum_{\mathbf{z}_j \in \mathcal{N}_5(\mathbf{z})}
        \sqrt{\sum_{i=1}^{3} (z_i - z_{j,i})^2}
\end{equation}

\textbf{Distance Threshold.}
The distance threshold $\tau$ is computed during offline calibration as the 99.9th percentile of all mean 5-NN distances observed across normal-training snapshots. Thus, $\tau$ is the largest distance that still covers 99.9\% of normal behavior. It is dimensionless, measured in Euclidean distance units of the standardized log-score (z-score) space. The remaining 0.1\% is deliberately excluded because it corresponds to noisy training snapshots with unusually low confidence scores across one or more experts. Including these outliers would expand the boundary and reduce sensitivity to genuine anomalies. In our training set, $\tau = 1.725$. A snapshot with $D \leq \tau$ is classified as normal, whereas a snapshot with $D > \tau$ is raised as an anomaly, regardless of whether any individual expert score crossed its per-channel threshold.

\textbf{PDP Trust Score.}
The distance is mapped to a continuous PDP trust score as
\begin{equation}
    P_{normality} = 0.5^{(D/\tau)} \times 100\%.
\end{equation}
This yields 100\% at $D{=}0$, 50\% at $D{=}\tau$, and exponentially decaying values for anomalies, enabling risk-proportional access control at the PDP.

\section{Experimental Evaluation} \label{sec:eval}

In this section, we present a multi-layered evaluation of our framework by evaluating each expert independently and then assessing end-to-end performance via the Multi-Expert Fusion Classifier. All experiments use Android \texttt{logcat} traces collected from a OnePlus~8T as the enrolled device. Normal-operation logs train the three expert models and calibrate the fusion classifier, while controlled perturbation datasets test identity OOV values, abnormal battery timing, unseen Wi-Fi environments, and benign Wi-Fi variations.

\subsection{Identity Expert Evaluation}
The Identity Expert's primary role is to validate the semantic integrity of dynamic variables within system logs. Unlike traditional anomaly detectors that treat variables as arbitrary strings, this expert evaluates whether a specific value (e.g., an IP address) is appropriate for its corresponding placeholder (e.g., \texttt{<ipv4>}) given the clean event-ID sequence and the previous dynamic-variable state of the Event ID under evaluation. OOV values and soft confidence deviations on known-vocabulary logs are assessed through the fusion stage.

\subsubsection{Ablation Study: The Impact of Sequence and Memory}
To quantify the contribution of contextual features, we performed an ablation study (Table~\ref{tab:IdentityAblation}), comparing the full model against a sequence-only variant and a context-free baseline. For each masked dynamic variable position, Top-1 accuracy measures whether the observed value is assigned the highest predicted probability by the model, whereas Top-3 accuracy measures whether the observed value appears among the three most probable candidates. The average confidence reports the mean Softmax probability assigned to the observed dynamic value.

\begin{table}[h!]
\centering
\caption{Ablation Study on Contextual Features}
\label{tab:IdentityAblation}
\begin{tabular}{@{}lccc@{}}
\toprule
\textbf{Configuration} & \textbf{Top-1 Acc.} & \textbf{Top-3 Acc.} & \textbf{Avg. Conf.} \\ \midrule
\textbf{SEQ + MEM (Full)} & \textbf{99.07\%} & \textbf{99.26\%} & \textbf{98.67\%} \\
SEQ Only                  & 23.37\%          & 38.26\%          & 21.32\%          \\
NONE (No Context)         & 14.21\%          & 25.13\%          & 14.07\%          \\ \bottomrule
\end{tabular}
\end{table}

The results demonstrate a substantial improvement when utilizing the stateful memory (\texttt{MEM}) feature. Given the clean 50-event-ID sequence and the previous values of the current Event ID, the model predicts newly masked values with 99\%+ Top-1/Top-3 accuracy, despite diverse  IPv4, IPv6, and MAC observations.

\subsubsection{Out-of-Vocabulary (OOV) Robustness}
During the OOV evaluation, the model was exposed to logs containing unknown MAC addresses, IPs, and SSIDs. These values are replaced by \texttt{[UNK]}, which was not observed in normal user-specific training contexts. The expert achieved a 100\% identification rate for OOV tokens, and the resulting confidence scores were naturally very low, substantially reducing $S_{Ident}$ and producing large outlier distances in the fusion stage.

\subsubsection{Value-Context Consistency Test}
To test whether the model truly understands placeholder semantics, we programmatically introduced value-context inconsistencies by assigning dynamic variable values, randomly, to incompatible placeholders (e.g., transplanting a MAC address or IP from one log into a different log's placeholder). As shown in Table~\ref{tab:IdentityValueContextConsistency}, the Overall Confidence
Score decreased from 98.23\% to 10.54\% as the inconsistency rate reached 100\%, suggesting that the model validates value-template compatibility rather than merely accepting known strings. Because \texttt{SEQ} contains only Event IDs, swapped values do not contaminate subsequent windows. Otherwise, a single injection could remain in context for 50 events and inflate false positives.

\begin{table}[h!]
\centering
\caption{Identity Expert Sensitivity to Value-Context Inconsistency}
\label{tab:IdentityValueContextConsistency}
\begin{tabular}{@{}cc@{}}
\toprule
\textbf{Inconsistency \%} & \textbf{Avg. Conf.} \\ \midrule
0\%   & 98.24\% \\
25\%  & 76.08\% \\
50\%  & 54.04\% \\
75\%  & 32.32\% \\
100\% & \textbf{10.54\%} \\ \bottomrule
\end{tabular}
\end{table}

\subsection{Battery Expert Evaluation}
The Battery Expert outputs a probability distribution over the six temporal classes; \texttt{TS\_VIOLATION} probability is folded directly into $S_{\mathrm{Batt}}$ as a logical-consistency penalty rather than a separate hard rule, so stochastic inter-event interval variability is absorbed by the learned nearest-neighbor normal-region manifold.

\subsubsection{Training Baseline}
To establish a reference for normality, we analyzed the global temporal characteristics of the training dataset (OnePlus~8T). Table~\ref{tab:TrainingDistribution} presents the Smoothed Dataset Distribution that defines the statistical baseline internalized by the BERT model. The baseline reveals that \texttt{DRAIN} events are concentrated almost entirely in \texttt{TS\_MID} (91.26\%), corresponding to a 1\% battery drop occurring every 9--29~minutes under typical usage. There are zero \texttt{DRAIN: TS\_INSTANT} instances in the normal set,
making rapid discharge a strong anomaly indicator in this dataset. \texttt{STABLE} events are dominated by \texttt{TS\_INSTANT} (57.20\%) and \texttt{TS\_SHORT} (41.08\%), reflecting the device's stable-state reporting behavior.

\begin{table}[h!]
\centering
\caption{Training Dataset Temporal Distribution}
\label{tab:TrainingDistribution}
\begin{tabular}{@{}lcc@{}}
\toprule
\textbf{Temporal Bin} & \textbf{STABLE Mode (\%)} & \textbf{DRAIN Mode (\%)} \\ \midrule
\texttt{TS\_INSTANT}   & 57.20 & 0.00           \\
\texttt{TS\_SHORT}     & 41.08 & 0.97           \\
\texttt{TS\_MID}       & 1.64  & \textbf{91.26} \\
\texttt{TS\_LONG}      & 0.08  & 7.77           \\
\texttt{TS\_SUSTAINED} & 0.00  & 0.00           \\ \bottomrule
\end{tabular}
\end{table}

\subsubsection{Abnormal Dataset Evaluation}
To evaluate the model's robustness, we generated an abnormal dataset subjected to a high-intensity resource-exhaustion attack. A stress-test application simultaneously activated the vibration motor, the LED flashlight, and maximum screen brightness while performing continuous CPU-bound computations.

This controlled stress significantly altered the device's battery-transition dynamics. Most importantly, physical battery depletion accelerated sharply: \texttt{DRAIN} events, corresponding to a 1\% battery-level drop, occurred every 1.5--2 minutes. This shifted the observed distribution from the normal \texttt{TS\_MID} region into \texttt{TS\_INSTANT} (83.33\%) and \texttt{TS\_SHORT} (16.67\%), as shown in Table~\ref{tab:AbnormalDistribution}. This change directly reflects the abnormal power-consumption pattern induced by the resource-exhaustion attack.

\begin{table}[h!]
\centering
\caption{Temporal Distribution During Resource-Exhaustion Attack}
\label{tab:AbnormalDistribution}
\begin{tabular}{@{}lcc@{}}
\toprule
\textbf{Temporal Bin} & \textbf{STABLE Mode (\%)} & \textbf{DRAIN Mode (\%)} \\ \midrule
\texttt{TS\_INSTANT}   & 100.00 & 83.33 \\
\texttt{TS\_SHORT}     &   0.00 & 16.67 \\
\texttt{TS\_MID}       &   0.00 &  0.00 \\
\texttt{TS\_LONG}      &   0.00 &  0.00 \\
\texttt{TS\_SUSTAINED} &   0.00 &  0.00 \\ \bottomrule
\end{tabular}
\end{table}

The Battery Expert achieved 100\% Recall on DRAIN events with zero false positives. Since the training distribution for \texttt{DRAIN: TS\_INSTANT} is 0.00\%, the model's predictive distribution remained centered on \texttt{TS\_MID}, \texttt{TS\_LONG}, and \texttt{TS\_SHORT}. The ground-truth \texttt{TS\_INSTANT} bin was assigned near-zero probability, substantially reducing $S_{Batt}$ and producing large outlier distances in the fusion stage on every affected event.

\subsection{Wi-Fi Topology Expert Evaluation}
The Wi-Fi Topology Expert is designed as a Dual-Head Reconstruction model. To evaluate its robustness against topological changes and Out-of-Distribution (OOD) environments, we conducted four stress-test scenarios (Table~\ref{tab:WifiScenarios}).

\begin{table}[h!]
\centering
\caption{Wi-Fi Topology Expert: Confidence Scores under Environmental Stress}
\label{tab:WifiScenarios}
\footnotesize
\begin{tabularx}{\columnwidth}{@{}>{\raggedright\arraybackslash}p{0.38\columnwidth}c>{\raggedright\arraybackslash}X@{}}
\toprule
\textbf{Scenario} & \textbf{Conf.} & \textbf{Analysis} \\ \midrule
1. Unknown Environment & \textbf{0.23\%} & Total OOD response. \\
2. Missing Access Points & \textbf{97.30\%} & Tolerates partial signal loss. \\
3. Signal Fluctuations ($\pm 1$) & \textbf{82.26\%} & Sensitive to local movement. \\
4. Signal Jump (Lvl 1$\rightarrow$4) & \textbf{71.60\%} & Detects significant spatial shifts. \\ \bottomrule
\end{tabularx}
\end{table}

The analysis reveals three key insights: (1)~the model is most sensitive to SSID mismatches as in Scenario~1, the confidence effectively vanished ($0.23\%$), indicating the model's sensitivity in identifying unfamiliar physical environments, (2)~the Self-Attention mechanism exploits topological redundancy, maintaining $97.30\%$ confidence even when 2 of~5~APs disappear, because the surviving APs maintain their expected spatial intensities and the model successfully reconstructs the known topology from partial context, and (3)~a minor signal fluctuation ($\pm 1$) is treated as expected environmental noise ($82.26\%$), while a major signal level jump (Scenario~4) is correctly flagged as a significant spatial shift ($71.60\%$), capturing sudden movements toward or away from individual routers.

\subsection{Multi-Expert Fusion Classifier Evaluation}
The final evaluation assesses the complete decision pipeline described in Section~\ref{sec:knn}. During inference, each incoming log event updates the persistent score snapshot $\mathbf{x} = [S_{Ident}, S_{Batt}, S_{WiFi}]$. The snapshot is transformed to log space, standardized using the normal-training statistics, and compared against the stored normal region using the mean 5-nearest-neighbor distance $D$. The calibrated 99.9th-percentile threshold was $\tau = 1.727$ and snapshots with $D > \tau$ are reported as anomalous and mapped to a PDP-facing normality score $P_{normality}$.

Table~\ref{tab:FusionSummary} summarizes the end-to-end behavior on normal operation and controlled perturbation datasets. These tests are designed to probe model sensitivity, not to represent complete real-world attack simulations. Each anomaly dataset is a mixed stream: injected events are labeled anomalous, while the remaining log entries retain normal labels. We report $\mathrm{FPR}=\mathrm{FP}/(\mathrm{FP}+\mathrm{TN})$ over normal-labeled samples and $\mathrm{FNR}=\mathrm{FN}/(\mathrm{FN}+\mathrm{TP})$ over anomalous samples. The normal dataset contains 2{,}041 log entries and produces 12 false positives, corresponding to an FPR of 0.59\%. The identity, battery, and new-environment Wi-Fi perturbations yield zero missed injected events in this controlled setting. Their nonzero FPR values are computed over the normal entries in the same mixed streams. The benign Wi-Fi perturbation tests contain no anomaly labels and therefore report only FPR.

\begin{table}[h!]
\centering
\caption{End-to-end fusion classifier performance on normal and injected-anomaly datasets}
\label{tab:FusionSummary}
\resizebox{\columnwidth}{!}{%
\begin{tabular}{lccccc}
\toprule
\textbf{Dataset} & \textbf{Samples} & \textbf{Detected} & \textbf{Missed} & \textbf{FPR} & \textbf{FNR} \\ \midrule
Normal operation              & 2041 & -- & -- & 0.59\% & -- \\
Identity OOV injection        & 2041 & 35 & 0 & 0.60\% & 0.00\% \\
Battery timing injection      & 2041 & 97 & 0 & 0.62\% & 0.00\% \\
New Wi-Fi environment         & 2041 & 50 & 0 & 0.45\% & 0.00\% \\
Minor RSSI fluctuations       & 2041 & -- & -- & 0.73\% & -- \\
Missing access points         & 2041 & -- & -- & 0.78\% & -- \\ \bottomrule
\end{tabular}%
}
\end{table}

These results indicate the intended role of the fusion layer. Strong controlled perturbations, such as OOV identity values, unseen battery-drain timing, or complete Wi-Fi topology changes, moved the snapshot outside the normal manifold in these datasets. At the same time, benign Wi-Fi changes do not trigger direct topology alerts, because the remaining known SSIDs and nearby RSSI levels preserve sufficient evidence of the learned environment. The residual false positives are dominated by the same borderline events observed in the clean normal trace, suggesting that errors arise from the global nearest-neighbor decision boundary rather than from the anomaly-injection procedure itself.

\section{Conclusion} \label{sec:conclusion}
This paper introduced a multi-expert BERT pipeline for Android-log-based continuous user-device authentication. The framework combines identity, battery, and Wi-Fi topology evidence and maps the fused score to a PDP-facing normality estimate. Controlled perturbation tests indicate that semantic, battery-timing, and topology deviations can be detected in the evaluated setting, while benign Wi-Fi variations remain largely tolerated with sub-1\% FPR. Future work will expand adaptive vocabulary updates, long-term drift handling, and broader multi-user validation.

\enlargethispage{3\baselineskip}
\makeatletter
\let\oldthebibliography\thebibliography
\let\endoldthebibliography\endthebibliography
\renewenvironment{thebibliography}[1]{\oldthebibliography{#1}\fontsize{6}{6.5}\selectfont}{\endoldthebibliography}
\makeatother
\bibliographystyle{IEEEtran}
\bibliography{references}

\end{document}